\documentclass{aastex631}
\usepackage{graphics}
\usepackage{enumitem}
\usepackage{amsfonts}
\usepackage{amsmath}
\usepackage{mathtools}
\usepackage{graphicx}
\usepackage{figsize}

\begin{document}

\title{Deciphering Pre-solar-storm Features Of September-2017 Storm From Global And Local Dynamics}

\author[0000-0002-0786-7307]{Breno Raphaldini}
\affiliation{High Altitude Observatory, NCAR \\
3080 Center Green Drive \\
Boulder, CO 80301, USA}

\author[0000-0002-2227-0488]{Mausumi Dikpati}
\affiliation{High Altitude Observatory, NCAR \\
3080 Center Green Drive \\
Boulder, CO 80301, USA}

\author[0000-0003-2622-7310]{Aimee A. Norton}
\affil{Hansen Experimental Physics Laboratory, 452 Lomita Mall, Stanford, CA 94305-4085, USA}

\author[0000-0001-5113-7937]{Andre S. W. Teruya}
\affiliation{Instituto de Astronomia, Geofísica e Ciências Atmosféricas\\ Universidade de São Paulo\\ São Paulo, Brazil}

\author[0000-0002-7369-1776]{Scott W. McIntosh}
\affiliation{High Altitude Observatory, NCAR \\
3080 Center Green Drive \\
Boulder, CO 80301, USA}

\author[0000-0002-0786-7307]{Christopher B. Prior}
\affiliation{Department of Mathematical Sciences, Durham University \\
Stockton Road
Durham-UK
DH1 3LE}

\author[0000-0003-2297-9312]{David MacTaggart}
\affiliation{School of Mathematics and Statistics University of Glasgow\\ Glasgow G12 8QQ, UK}

\begin{abstract}
We investigate whether global toroid patterns and the local magnetic field topology of solar active region AR12673 together can hindcast occurrence of the biggest X-flare of solar cycle (SC)-24. Magnetic toroid patterns (narrow latitude-belts warped in longitude, in which active regions are tightly bound) derived from surface distributions of active regions, prior/during AR12673 emergence, reveal that the portions of the South-toroid containing AR12673 was not tipped-away from its north-toroid counterpart at that longitude, unlike the 2003 Halloween storms scenario. During the minimum-phase there were too few emergences to determine multi-mode longitudinal toroid patterns. A new emergence within AR12673 produced a complex/non-potential structure, which led to rapid build-up of helicity/winding that triggered the biggest X-flare of SC-24, suggesting that this minimum-phase storm can be anticipated several hours before its occurrence. However, global patterns and local dynamics for a peak-phase storm, such as that from AR11263, behaved like 2003 Halloween storms, producing the third biggest X-flare of SC-24. AR11263 was present at the longitude where the North/South toroids tipped-away from each other. While global toroid patterns indicate that pre-storm features can be forecast with a lead-time of a few months, its application on observational data can be complicated by complex interactions with turbulent flows. Complex/non-potential field structure development hours before the storm are necessary for short term prediction. We infer that minimum-phase storms cannot be forecast accurately more than a few hours ahead, while flare-prone active regions in peak-phase may be anticipated much earlier, possibly months ahead from global toroid patterns.

\end{abstract}


\section{Introduction}\label{sec:intro}

Understanding and ultimately predicting intense explosive events such as large flares and coronal mass ejection (CME) events are currently the cornerstones of solar research due to their socio-economic-enviromental impacts \citep{khodairy2020impact,vanselow2020solar,simpson2011possibility}. One approach that has been explored in order to understand the circumstances that lead to large flares and CMEs is to study local physical properties in the vicinity of active regions such as magnetic fluxes \citep{jing2010free,metcalf2005magnetic, regnier2007free}, plasma velocity fields \citep{attie2018}, magnetic helicity \citep{pariat2005photospheric,demoulin2009modelling,gupta2021magnetic}, magnetic winding \citep{mactaggart2021direct,raphaldini2022magnetic} and magnetic twist \citep{kusano2020physics}. The evaluation of the predictive skill about several quantities and potential operational applications can be found in a series of papers \citep{barnes2016comparison, leka2019comparison, leka2019comparison2,park2020comparison}. Studies of plasma flows in flux emergence simulations on the other hand show that it is possible to predict new emergences within 10 hours of advance \cite{silva2023novel}. On the other hand, the global pattern of the toroid, the strong reservoir of magnetic field stored at depth in the Sun and from which the active regions emerge, can provide certain features indicative of the occurrence of big solar storms \citep{dikpati2021deciphering}.

The spatio-temporal organization of solar active regions follow well-known patterns both in latitude and longitude. The latitudinal migration is exemplified by the so called butterfly-diagram, from mid-latitudes ($\sim 35^{\circ}-30^{\circ}$) towards the equatorial region as the solar cycle progresses \citep{hathaway2015solar}. The butterfly diagram is understood in terms of the migration of toroidal magnetic field bands following the dynamo cycle and has been simulated using various dynamo models \citep{dikpati2009flux,rudiger1995solar,cameron2017global}. Less obvious is the organization of active regions in longitude, namely the activity-nests and long-lived preferred longitudes that can persist for long times, possibly beyond one solar cycle \citep{neugebauer2000solar,balthasar1983preferred}. After the discovery of solar Rossby waves from observations \citep{mcintosh2017,loptien2018global},  preferred longitudes, where the solar active regions' manifestations recur, have been explored by simulating the interactions among Rossby waves, differential rotation and toroidal magnetic fields. The tachocline is a likely region for these nonlinear interactions to take place, and sustained bulging at certain latitude-longitude locations are favorable for flux-emergence and can plausibly be the cause of active regions' nesting \citep{dikpati2020space}.

Rossby waves are large-scale propagating patterns of vertical vorticity long known to exist in Earth's atmosphere and oceans \citep{Rossby1939,madden1979observations}, and are one of the key physical mechanisms for understanding of global weather and climate dynamics \citep{hoskins1990, hoskins2002, hoskins1993}. While Rossby waves have long been hypothesized to exist in the Sun \citep{gilman1968}, the relationship between the dynamics of Rossby waves, modified by the presence of magnetic fields, and the longitude-dependent magnetic activity of the Sun was explored only recently \citep{dikpati2020space}.

Rossby waves in the Sun, from observations and theory, can be classified to two distinct groups, namely those that are dynamic, and hence can shape the magnetic fields and patterns, and those that are of diagnostic value but for which individual modes are two low in amplitude to have influence on magnetic patterns. The dynamic ones are of relatively low longitudinal wave number ($m=1$, 2 and 3) and there are only a few modes, each of which has significant amplitude. These are mostly generated from global MHD instabilities \citep{dziembowski1987low,gilman1997joint,Zaqarashvili_2010b,raphaldini2015nonlinear,Dikpati_2018} . Their signatures are observed in the solar magnetic patterns at the solar surface, in the spatial distribution of active regions \citep{dikpati2021deciphering}, and at the corona, in the evolution of bright point densities \citep{mcintosh2017}, and in the longitude-drift of long-lived coronal holes \citep{Harris_2022}. It is to be noted that there can be other Rossby waves which can interact with mean magnetic fields. However, the Rossby waves that can interact locally with the dynamo-generated toroidal magnetic fields will have to be located at/near the base of the convection zone.

\citet{Korsós_2023} provided evidence for the link between intermediate period, associated with meridional oscillations of photospheric structures and the occurrence of large flares, suggesting that magnetic Rossby waves play a major role in the occurrence of these events. By contrast, the helioseismically determined Rossby waves are many in numbers \citep{loptien2018global}  and each mode is of low in amplitude. They include other inertial oscillations too \citep{hathaway2022variations}. These Rossby waves can be understood from simulations also near the surface by inverse-cascade of kinetic energy, such as horizontal supergranular motions \citep{dikpati2022simulating}. Most recently theory has shown the possible existence of thermal Rossby waves, whose properties are tied to the outward density-decline in the convection zone coupled with rotation and differential rotation \citep{hindman2022radial}. They are yet to be observed. As more becomes known about this class of Rossby waves, the concept presented in this paper may need modification. But at the present time, to understand the pre-solar-storm properties from global magnetic patterns, we focus on the Rossby waves generated by global MHD instabilities, and elaborate on that in the following paragraph.

Studying properties such as characteristic frequencies and growth rates, \citep{Zaqarashvili_2010,Zaqarashvili_2010b} suggested that magnetized Rossby waves in the solar tachocline were plausible sources for intermediate frequencies of solar magnetic activity (from several months up to a few years) such as Rieger-type periodicities and the quasi-biennial oscillation. Using a nonlinear magnetohydrodynamic shallow water model, \citet{Dikpati_2018} showed that there is a good match between the characteristic phase speed of the magnetized Rossby waves and the longitudinal migration of magnetic field structures observed in the photosphere. \citep{Dikpati_2017,Dikpati_2018seasons} also showed, within the same framework, that an oscillatory exchange of energies among Rossby waves, differential rotation and magnetic fields gives rise to a so-called Tachocline Nonlinear Oscillation (TNOs), which can explain bursty ``seasons'' of solar activity in  intermediate timescales of a few weeks to several months. A similar type of nonlinear interaction between Rossby waves and zonal structures have also been invoked as a mechanism for longer term periodicities of the Sun including the occurrence of Grand minima \citep{raphaldini2015nonlinear,raphaldini2019,Raphaldini_2020}.

Because the subject of solar Rossby waves is relatively new, at this point it is worthwhile to discuss analogies between the roles played by atmospheric Rossby waves on weather and climate and that played by solar Rossby waves on the magnetic activity of the Sun. It is well-established that atmospheric Rossby waves constitute a mechanism that organizes weather patterns on large scales. The so-called teleconnection patterns, consisting of geographically widely separated areas that have connected weather oscillations \citep{hoskins1990}, such as observed in rainfall time-series \citep{boers2019}. The link between Rossby waves and global rainfall patterns lie in the mechanisms of formation of clouds and their relationship with low/high pressure regions due to Rossby waves. Solar Rossby waves in the tachocline are also expected to generate sequences of high and low pressure regions, in the formation of bulges in the top of the tachocline \citep{Dikpati_2018}. These bulging patterns will determine locations from where magnetic fields will emerge through the convection zone, eventually reaching the photosphere and giving rise to active regions. 

The association between atmospheric Rossby waves and weather systems was introduced by \citet{sutcliffe1947contribution} and resulted in a substantial advance in the weather prediction practices before the era of super-computing. Monitoring Rossby waves' troughs and ridges (high/low pressure regions) became an operational practice that expanded the prediction horizon from a few hours to several days  compared with purely local observations of the cloud conditions \citep{bluestein1992synoptic}. With this solar-terrestrial weather analogy in mind, can monitoring solar Rossby waves extend the prediction horizon from a few hours to several days/weeks compared with pure observations of active regions' physical properties? 

Having discussed the important roles a class of solar Rossby waves plays in organizing longitude distribution of active regions (see also \citep{raphaldini2023information,teruya2022ray}, analogous to the way atmospheric Rossby waves do for terrestrial weather patterns, we explore pre-storm feature analysis from solar Rossby waves' dynamics. \citet{dikpati2021deciphering} derived toroid patterns by fitting low-order longitudinal modes to synoptic magnetogram maps of magnetogram. When applied to a pre-storm configuration, namely the ``Halloween Storms" of 2003, originated from a X-45 class flare \citep{thomson2004ionosphere}, the wavy toroid patterns in the north and south hemisphere were found to tip away from each other at certain longitudes, and consequently at certain other longitudes the north and south toroids were found to come closer to each other. It was argued that the "tipped-away" portions of the toroids at a specific longitude led to big storms, first from the south and then in 2-3 days from the north. The results were interpreted in terms of bulging generated by shallow-water Rossby waves and their perturbation to the toroidal magnetic field at the solar tachocline. Those simulations showed how fluid-bulges' movement from the south to the north in a couple of days can create "imprints" of big spot emergences from the same longitude in both hemispheres in the interval of a few days.

Big storms of similar strength to the Halloween storms did not occur in cycle 24, which was the weakest cycle in 100 years. Instead, the biggest storm occurred on September 6, 2017, which was essentially in the minimum-phase of that cycle. We seek to answer the following questions in this study: (i) can the global toroid patterns be derived for a storm that occurred during the minimum-phase of a cycle, when there are very few spots? (ii) What can be inferred about the pre-storm features in 2017 from those toroids? (iii) Even though rare, if the storms occur in solar minimum-phase, and hence the global patterns may lack pre-storm features from which to make forecasts, what can be inferred instead from local dynamics of the relevant active regions, in this case AR12673? (iv) What are the differences in pre-storm features derived from global and local patterns when the storms occur in a maximum-phase compared to a minimum-phase?
 
If global toroid patterns can indicate pre-storm features, a lead time of a few weeks can be available to prepare for the prevention of hazardous space weather impacts. However, global toroid patterns can not be very meaningfully constructed when there is only one spot, such as that often happens during very late declining phase of a cycle or during solar minimum. In those cases, if a big X-class flare occurs, local dynamics would be the only option to derive pre-storm features, but we lose the weeks longer lead-time, because the indication from local dynamics provides only several hours lead-time. It is clear that to best utilize the predictive capability of models that can provide pre-storm features, we need to study both the global patterns of relevant active regions with a few weeks' lead-time as well as their local dynamics with a few hours' lead-time. In the present paper, we aim to study the connection between large-scale deformation patterns observed in the latitude-longitude organization of active regions observed from magnetic field synoptic maps, and the complexity of the active regions that led to the two largest flares during solar cycle 24. The manuscript is organized as follows: in section (2) we describe the methods used: (i) analysis of global magnetic field patterns through toroid fitting and (ii) local analysis via topological characterizations of active regions via magnetic helicity and magnetic winding analyses. In section (3) we describe the results from the global and local analysis both in a off-maximum storm (AR 12673) and for an active region in the solar maximum (AR 11263) which is also compared with an adjacent non-flaring region (AR 11266). Section (4) gives a summary, and section (5) presents conclusions with accompanying discussion.

\section{Methods}

As discussed in \S1, we explore the pre-solar-storm features from global and local dynamics, to understand how far ahead of time the occurrence of the storm can be estimated. We discuss briefly in \S2.1 the methodology for deriving the evolution of the global toroid, from which surface active regions manifest, by implementing the "Trust Region Reflective" algorithm (see, e.g, \citet{dikpati2021deciphering}, for details). In \S2.2 we describe the procedure for calculating helicity and winding patterns for the active regions of our interest (see \citet{raphaldini2022magnetic}). We select AR12673, which caused the biggest X-class flare of cycle 24 during September 2017. Noting that the biggest X-flare occurred in the late declining phase of cycle 24, almost during the minimum phase, we also select another active region, AR11263, which caused another big X-class flare during 2011, during the late rising phase just before the peak of cycle 24. Thus we can compare how the global and local dynamics can be indicative of pre-storm properties at solar minimum and maximum phases. 

\subsection{Deriving toroid patterns of distribution active regions}

Here we follow the formulation introduced in detail in \citet{dikpati2021deciphering} for fitting toroid patterns to magnetic field synoptic maps. In the present study we use HMI synoptic maps, which are constructed by combining near central meridian data from 20 magnetograms, resulting in a 1440 $\times$ 3600 gridpoint sine(latitude) $\times$ longitude charts. These maps provide a representation of the latitudinal distribution of active regions as well as their Carrington longitude positions. The main idea introduced in \cite{dikpati2021deciphering} is to represent the spatial distribution of active regions in terms of two wavy-belts or toroids, one for each hemisphere, constructed by a superposition of Fourier modes.
Given a synoptic chart at time $t$, global toroid patterns $P_c(\phi,t)$ are determined via a superposition of Fourier modes of the form:
\begin{equation}\label{Fourier}
    P_c(\phi,t)=\sum_{m=0}^N q_m(t) sin(m\phi + \zeta_m(t))
\end{equation}

Where $q_m(t)$ is the Fourier amplitude of the m-th mode at time $t$ and $\zeta_m(t)$ the respective Fourier phase. Both Fourier amplitude and phases are represented as a combination of mean and time-varying parts
\begin{align}
    q_m(t)=\overline{q}_m+\sum_i s_{i,m}(t_i)\\
    \zeta_m(t)=\overline{\zeta}_m+\sum_i w_{i,m}(t_i)
\end{align}

where $s_{i,m}$ and $w_{i,m}$ represent, respectively, the time-varying parts of amplitudes and phases at time $t_i$, with $i$ representing a discrete time index.
In (\ref{Fourier}), $q_0$ represents the mean latitudinal position of the active region configuration at time $t$. An optimization scheme is then employed to constrain the number of parameters used to represent the toroid belt configuration. The "Trust Region Reflective" (TRR) algorithm, introduced in \citet{branch1999subspace}, is a nonlinear  optimization algorithm that consists of replacing a function over a high dimensional functional space by a simpler function that reasonably represents the objective function in a region of the parameter space called the ``trust region", then the variance of the approximate function is iteratively minimized over the trust region until a local minima is reached. Once a local minimum is reached the algorithm expands the trust region over which it searches for a minimum. In our study, Trust Region Reflective (TRR) algorithm converges successfully for longitudinal modes up to $m=10$ or so, which is more than sufficient for deriving toroid patterns of our interest, given that we do not see more than 10 latitude-longitude locations where active regions appear. We note that the sequence of times at which the synoptic maps are available is discrete, one for each Carrington rotation, representing a non-sychronous snapshot of the magnetic field configuration of the Sun as observed from the central meridian. Consequently, the time at which significant flare activity takes place will not necessarily agree with the time at which the active region pass through the central meridian. In the meantime, the active region may slightly move in latitude. Therefore, it is important to consider not only the Carrington rotation at which the flare activity took place, but also previous and following Carrington rotations.

\subsection{Topological measures of complexity of active regions}

Magnetic helicity has been extensively used in plasma physics, and particularly in the context of solar physics. It is a conserved quantity in ideal magnetohydrodynamics, and it is generally well preserved even if diffusive effects are present, with slow dissipation rates \citep{biskamp2003magnetohydrodynamic}. A version of the magnetic helicity that is particularly useful for solar magnetic fields is the relative helicity, the helicity calculated relative to a basic state consisting of a potential magnetic field. This version of the magnetic helicity can be computed solely from photospheric fluxes. Early examples of the computation of magnetic helicity budgets in active regions can be found in \citet{chae2001observational,chae2004determination,kusano2002measurement,kusano2004trigger} including the suggestion of possible relationships with flare activity \citep{moon2002flare}. 

Several studies point to the relative importance of current-carrying helicity for the flaring activity. The idea is simple, since potential magnetic fields do not dissipate energy, the more non-potential a magnetic field configuration is, the more available energy there is to be dissipated in the form of flares. \citet{pariat2017} proposed, based on numerical simulations, that the ratio between the current-carrying and total relative helicity should be a good indicator of the occurrence of intense flares. The idea of measuring the relative importance of the current-carrying helicity has been applied to the study of observations from active regions that resulted in large flares  \citep{James_2018,Green_2022, moraitis2019magnetic}, confirming that a higher degree of current-carrying helicity preceeds large flares. A recent study \citep{Liu_2023} performed an extensive study, comprising 21 X-class flares that occurred during solar cycle 24,  showing important changes in the free magnetic energy and helicity in the corona, as well as their photospheric fluxes.

More recently, a quantity related to the magnetic helicity, called magnetic winding, was introduced in the study of solar magnetic fields \citep{mactaggart2021helicity,mactaggart2021direct,prior2020,raphaldini2022magnetic}. The magnetic winding is related to a well-known topological quantity called linking number \citep{arnold2008}, and can be seen as a measure of entanglement between field lines. Similar to the studies showing that a high contribution of current carrying helicity is indicative of large flare activity, \citet{raphaldini2022magnetic} showed that magnetic field configurations dominated by current-carrying magnetic winding can also be an indicator of strong flare activity. 

In the present study we use magnetic helicity and winding as measures of the complexity of the magnetic field configurations in active regions in Carrington rotations presenting strong flares. More specifically, we will study the two active regions that produced the strongest flares during solar cycle 24: AR12673, which produced the two largest flares of that cycle (one X-9 and one X-8 flare), and AR11263, which produced the third largest flare of that cycle (an X-6 flare). Here we briefly  describe the techniques used to compute the magnetic helicity and winding; further details can be found in \citet{raphaldini2022magnetic}.

\subsection{Magnetic Helicity}

We start from the definition of relative helicity \citep{berger1984topological} on a bounded volume V with a smooth boundary $\partial V$:

\begin{equation}\label{relhel}
    \mathcal{H}_R=\int_V (\textbf{A}+\textbf{A}_P).(\textbf{B}-\textbf{B}_P)\,{\rm d}^3 x
\end{equation}

Where $\textbf{B}_p$ is a reference potential field with a normal component at the surface $\partial V$ coinciding with that of $\textbf{B}$, $\textbf{B}\cdot\textbf{n}=\textbf{B}_p\cdot\textbf{n}$, where $\textbf{n}$ is the unit vector pointing outward at the surface $\partial V$. Here we place the surface boundary $\partial V$ placed at the photosphere, while the volume $V$ is located at the solar atmosphere.$\textbf{A}$ and $\textbf{A}_P$ are respective vector potentials (i.e. $\nabla \times \textbf{A}= \textbf{B}$ and $\nabla \times \textbf{A}_P= \textbf{B}_P$) which are defined up to an irrotational gauge that does not change the value of the integral in (\ref{relhel}). The calculation of (\ref{relhel}) requires the knowledge of the magnetic field in the whole volume $\partial V$, in the study of magnetic fields in active regions, this would imply in the knowledge of the magnetic field in the chromosphere. At this point an assumption has to be made: one alternative is to reconstruct $\textbf{B}$ from the  knowledge its value on the surface $\partial V$; another  common procedure is to assume that $\textbf{B}$ is a force-free field, and then reconstruct it from the magnetic field components on $\partial V$ \citep{Thalmann_2012,Thalmann_2011}. An alternative approach is to calculate the helicity fluxes though the photosphere. By integrating the fluxes in time one can obtain the accumulated  helicity thoughout the passage of the active region on the disk \citep{Romano_2014,mactaggart2021direct,raphaldini2022magnetic}. Here, we will follow the second approach. 
The helicity flux though the surface $\partial V$ is computed as:

\begin{equation}\label{hel1}
    \frac{d {H}}{dt}=\int_{\partial V}\frac{d \mathcal{H}(\textbf{x})}{dt}{\rm d}^2 x=-\frac{1}{2\pi}\int_{\partial V}\int_{\partial V}B_z(\textbf{x})B_z(\textbf{y})\frac{(\textbf{u(x)}-\textbf{u(y)})\times(\textbf{x}-\textbf{y})}{\vert \textbf{x}-\textbf{y}\vert ^2} {\rm d}^2 y {\rm d}^2 x
\end{equation}
Here $\textbf{u}(\textbf{x})$ is the velocity of the foot-point associated with the magnetic field line at a point $\textbf{x}$ in the photospheric plane. $\mathcal{H(\textbf{x})}$ is the so called field-line helicity, which corresponds to the helicity density at point $\textbf{x}$.

Consequently the accumulated helicity between times $0$ and $T$ is computed by integrating (\ref{hel1}) in time:

\begin{equation}
    H=-\frac{1}{2\pi}\int_0^T\int_{\partial V}\int_{\partial V}B_z(\textbf{x})B_z(\textbf{y})\frac{(\textbf{u(x)}-\textbf{u(y)})\times(\textbf{x}-\textbf{y})}{\vert \textbf{x}-\textbf{y}\vert ^2} {\rm d}^2\textbf{y} {\rm d}^2\textbf{x}
\end{equation}

Details on the potential/current-carrying decompositions of the magnetic helicity are found in the Appendix, also see \citet{raphaldini2022magnetic} for further details.

\subsection{Magnetic winding}
Similarly, the magnetic winding can be computed via a renormalization of the magnetic field to a unit value:
\begin{equation}
 f(x)= 
\begin{dcases}
    I(\textbf{x})=1 \quad if \quad B_z(\textbf{x})>0\\
    I(\textbf{x})=0 \quad if \quad B_z(\textbf{x})=0\\
    I(\textbf{x})=-1 \quad if \quad B_z(\textbf{x})<0
\end{dcases}
\end{equation}

then the magnetic winding flux is defined as:

\begin{equation}
    \frac{d {L}}{dt}=-\frac{1}{2\pi}\int_{\partial V}\int_{\partial V}I(\textbf{x})I(\textbf{y})\frac{(\textbf{u(x)}-\textbf{u(y)})\times(\textbf{x}-\textbf{y})}{\vert\textbf{x}-\textbf{y}\vert ^2} {\rm d}^2\textbf{y} {\rm d}^2\textbf{x}
\end{equation}
and the accumulated magnetic winding in the time interval $0$ and $T$:

\begin{equation}
    L=-\frac{1}{2\pi}\int_0^T\int_{\partial V}\int_{\partial V}I(\textbf{x})I(\textbf{y})\frac{(\textbf{u(x)}-\textbf{u(y)})\times(\textbf{x}-\textbf{y})}{\vert \textbf{x}-\textbf{y}\vert ^2} {\rm d}^2\textbf{y} {\rm d}^2\textbf{x}dt
\end{equation}

Respective components of the magnetic winding, namely its potential and current-carrying components, are defined in a completely analogous way to the definitions of helicity decompositions and so is omitted here; for further details see \citet{raphaldini2022magnetic}. The winding can  be seen as a renormalization of the magnetic helicity, by replacing the magnetic field by a field with the exact same orientation, but unit intensity. This can emphasize certain regions of the active region domain with complex structures, but weak radial field, such as polarity inversion lines \citep{mactaggart2021direct,raphaldini2022magnetic}.

\subsection{Measures of current-carrying/potentiality}
As previously discussed, there is a substantial literature pointing out the role of emergence of current-carrying dominated structures before substantial flare activity \citep{pariat2017,Green_2022,raphaldini2022magnetic}. Potential magnetic fields cannot dissipate energy through ohmic processes, but the emergence of current-carrying structures can lead to substantial release of energy in the form of flares. 

\citet{raphaldini2022magnetic} introduced a measure of the imbalance between current carrying and potential components of magnetic helicity and magnetic winding, named $\delta$ measures. First, for the magnetic helicity, this quantity measures at each point of the domain $\textbf{x}$ the difference between the contributions of the current carrying component of the helicity density $\mathcal{H}_c(\textbf{x})$ and the potential component of the helicity $\mathcal{H}_p(\textbf{x})$. We define this instantaneous imbalance as:

\begin{equation}\label{deltahel}
    \delta H'=\int_P \big( \vert \mathcal{H}_c(\textbf{x}) \vert-\vert \mathcal{H}_p(\textbf{x}) \vert\big){\rm d}^2\textbf{x}
\end{equation}

Analogously the delta measure is defined for the winding as

\begin{equation}\label{deltawin}
    \delta L'=\int_P \big( \vert \mathcal{L}_c(\textbf{x}) \vert-\vert \mathcal{L}_p(\textbf{x})\vert \big){\rm d}^2\textbf{x}
\end{equation}

These quantities are usually very noisy; in the analysis performed in the Results section we utilize a smoothed version of these quantities over a time-window of 10 hours.

\subsection{Helicity/winding computations}

Here, the computed quantities associated with the magnetic helicity and magnetic winding are performed using the ARTop code \footnote{https://github.com/DavidMacT/ARTop} \citep{alielden2023artop}. This code computes helicity and winding quantities using the Space-Weather HMI Active Region Patches (SHARP) datasets corresponding to 12-minute cadence magnetograms for a given active region. 
The code then used the Differential Affine Velocity Estimator for Vector Magnetograms (DAVE4VM) method \citep{schuck2008tracking} to obtain the velocity fields. DAVE4VM uses sequential vector magnetogram images, and assumes a locally linear (or affine) approximation of the velocity field, the evolution of the velocity field is obtained by fitting the linear (affine) coefficients  from the evolution of the magnetic field vectors.

\section{Results}

\subsection{A Solar Cycle Minima-phase Storm}

Active region AR 12673 is well known for being very flare prolific, producing four X-class flares and 27 M-class flares, including the two largest flares of SC-24 (an X-9.3 and an X-8.2 flares) \citep{sun2017super}. The study of this active region, therefore, constitutes a testbed for understanding the circumstances under which flare-prolific active regions are formed, in terms of both global and local dynamics.

\begin{figure}[ht]
    \centering
    \includegraphics[width=0.85\linewidth]{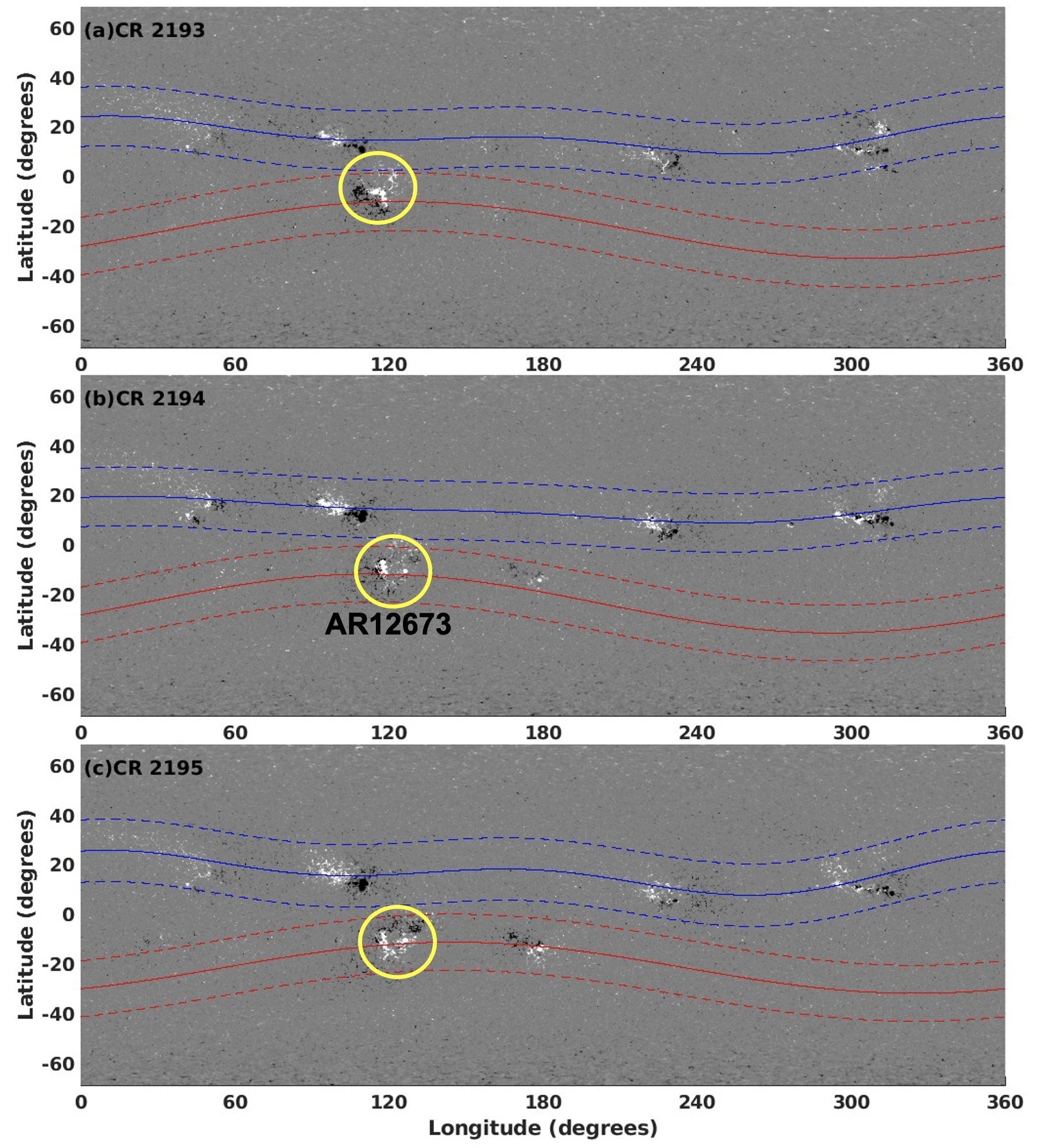}
    \caption{Evolution of magnetic toroid pattern during Carrington Rotations CR2193 (Jul 20-Aug 16, 2017), CR2194 (Aug 16- Sep 12, 2017) and CR2195 (Sep 12- Oct 10,2017), i.e., from one CR before through the storm to one CR after, during which AR12673, located at 120-degree longitude in South (yellow-circled region), produce two biggest flares (X9.3 and X8.2) of cycle 24, during late declining phase near solar minimum.
\label{Minima-phase-storm}    
    }
\end{figure}

One possible reason for the remarkable flare activity of AR 12673 is that it resulted from an emergence at the same location as an old decaying spot (AR 12665 and 12670). Additionally, AR 12673 was characterized by a very large magnetic flux emergence rate \citep{sun2017super}, as well as the strongest magnetic field ever recorded in the Solar corona. We will investigate the reasons why AR 12673 resulted in such strong flare activity, by analysing the global magnetic field morphology during Carrington rotations 2193-2195 in subsection \ref{12673global}. In subsection \ref{12673local} that follows, we analyse the magnetic helicity and magnetic winding contained in AR 12673, with particular emphasis on the role of current-carrying/potential-field decompositions.

\subsubsection{Analyses of pre-storm features of AR12673 from global toroid pattern}\label{12673global}

Global toroid patterns of emerged active regions were previously shown to infer certain pre-storm features, namely for the Halloween storms of 2003 \citep{dikpati2021deciphering}. Primarily three active regions ARs 10484 and 10488 in North and AR 10486 in South created the biggest solar storms of cycle 23, causing 11 X-class flares and 46 M-class flares, the largest being a X-28 flare, during a short span of time from October 19 to November 5 of 2003. While AR 10486 was the biggest active region of cycle 23, and was a complex $\beta\gamma\delta$-type, much was learned by analyzing the toroid patterns a few CRs before the Halloween storms of CR2009. Figures 10 and 11 of \citet{dikpati2021deciphering} show how the sustained dominant $m=2$ longitudinal mode, which produced longitude regions of closest proximity between the North and South toroids as well as furthest-away points respectively, created weakening and strengthening of magnetic fields. The furthest-away points at the same longitude in the North and South toroids can be created from the antisymmetric 'tipping' of toroids at their deeper origin, at or near the base of the convection zone or tachocline, due to global MHD instability of spot-producing toroidal magnetic bands. The nonlinear evolution of such instability produced patterns, being governed by the interaction of Rossby waves with toroidal band, can be sustained for a few CRs, because Rossby waves drift slowly in longitude. Thus from such nonlinearly evolving toroid patterns the upcoming Halloween storms of 2003 were hindcast and also simulated using a global MHD shallow-water model (see, e.g., figures 12, 13 and 14 of \citet{dikpati2021deciphering}) .

To examine whether we can analyse similarly the pre-storm features of the biggest storm of cycle 24, which occurred during 2017 September, we derive the toroid patterns for CRs 2193-2195, namely from one CR before the occurrence of the storm through one CR after it. Figure \ref{Minima-phase-storm} displays how the North and South toroids evolved. Three panels of synoptic maps clearly reveal several interesting features. First, the toroid patterns are dominated by the $m=1$ mode in longitude, even though several higher longitudinal wave numbers are included during the optimization procedure. This is not surprising, because during a late declining or solar minima phase, only a few active regions may emerge. Mode-fitting in such cases should favor the $m=1$ mode to be the dominant one. We see that there is one strong and two weak active regions in the North and one moderate active region in the South. 

Second, we find that North and South toroids are locally tipped in anti-phase about the equator, but unlike the Halloween storm of cycle 23, the biggest flare-producing active region AR12673 of cycle 24 is not located in the tipped-away portion of the toroid. Instead it is in close proximity of to another active region of the South toroid. So, following a similar logic given for Halloween storms' occurrence at the same longitude from tipped-away portion of the North and South toroids, we note that the global dynamical interactions of North and South toroids would indicate weakening in their field-strengths, and hence hindcasting pre-storm properties accurately with a lead time of at least one CR before the storm, is not possible in this case; perhaps this is true in the case of majority of solar minima-phase storms.

In fact, weakening of the active regions when the North and South toroids are in close proximity at a certain longitude was really the case for AR12673 initially; it was decaying and became an apparently inactive spot, and then suddenly got converted to the most active complex region after a new emergence occurred at the center of AR12673, leading to the biggest flare of that cycle. The new emergence at the center of the old, decaying active region, as well as complex interactions among the North and South toroids' ARs at the same longitude, produced a huge build-up of magnetic helicity. Thus, the local dynamics must have played a crucial role a few hours before the trigger of the storm, despite the global dynamics not anticipating a big storm a CR before. In this case, The global toroid patterns derived from the surface distribution of active regions may not be the best indicator of the dynamics of the spot-producing toroidal fields at the bottom, because only very few flux emergences occur then. The 2017 storm is an unusual storm, which has been difficult to understand, simulate and predict..
\begin{figure}[ht]
    \centering
    \includegraphics[width=0.8\linewidth]{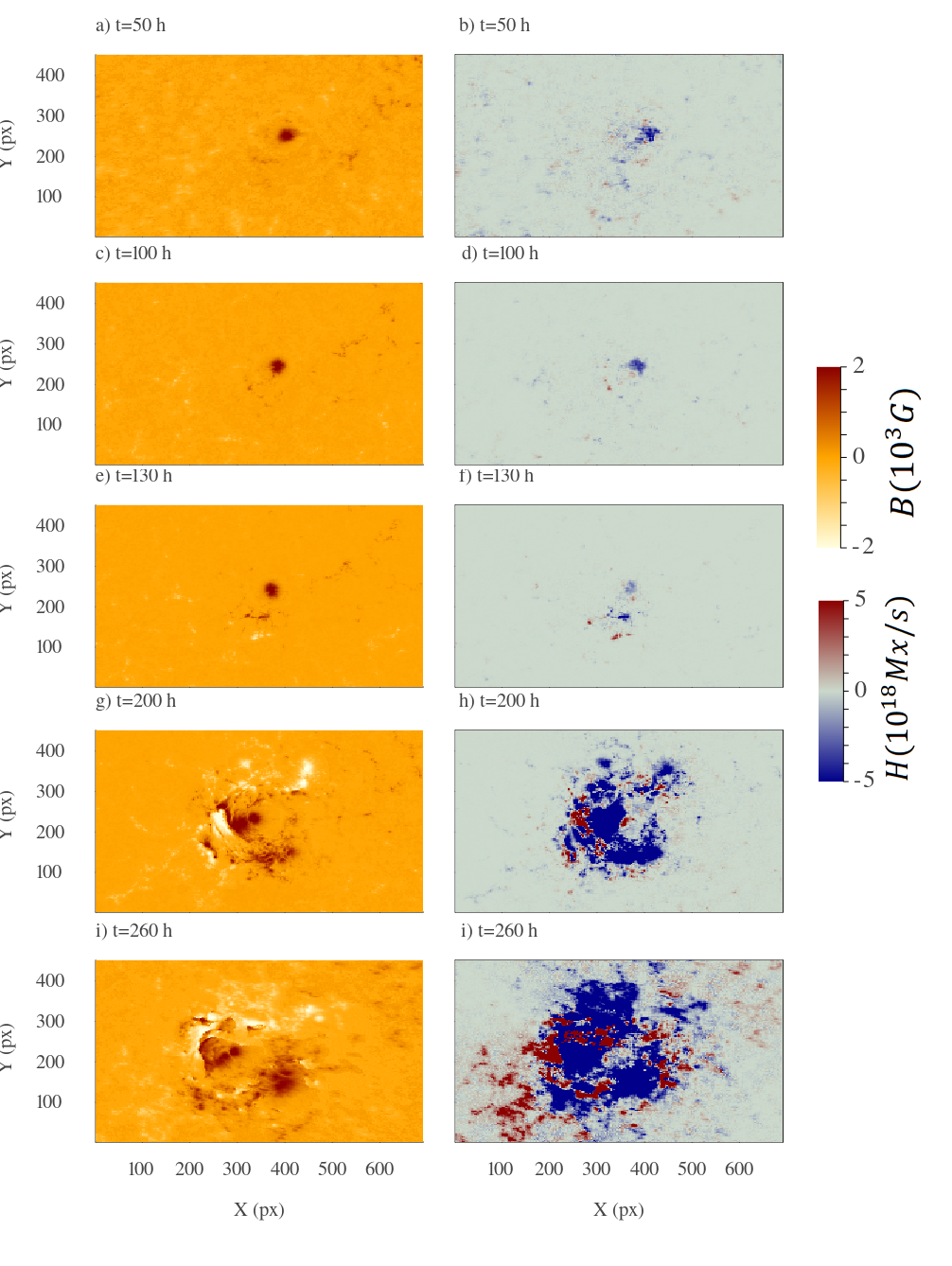}
    \caption{Evolution of the magnetic fields (left) and helicity  flux (right) for AR 12673. Time evolution movies of AR 12673 can be seen \href{https://zenodo.org/record/8305952}{in continuum intensity and magnetogram. The dimensions of the pixels are defined in \citep{bobra2014helioseismic}}}.
    \label{maghel}
\end{figure}
Local dynamics itself leads to only a short lead time of a few to several hours; nevertheless we need to include knowledge gained from local dynamics in pre-storm predictive feature analysis for all storms, but particularly for solar minimum storms. When we examine the build-up of helicity and winding, which exceeds a certain high value to trigger the flares, we point out an important feature derived from the global dynamics for this specific storm during solar minimum-phase. In particular, how was there a new emergence at the center of the old, decaying spot, which eventually became the most active spot of that cycle? As conjectured and demonstrated by \citet{dikpati2020space}, "imprints" of flux emergence correlate with the latitude-longitude location of tachocline bulges, which push up the toroidal field at that latitude-longitude location into the convection zone for their eventual appearance at the surface (see, e.g., the figure 12 of \citet{dikpati2020space}). Such bulges, if sustained for several CRs, depending on the global nonlinear dynamical interaction among the differential rotation, toroidal fields and Rossby waves, can lead to multiple emergences at the same latitude-longitude. This is what happened for the new emergence at the center of the old decaying AR12673. To understand how that new emergence lead to complex interactions with the old inactive AR to make it the most active one of cycle 24 we consider the role of local dynamics in the next subsection.

\subsubsection{Analyses of pre-storm features of AR12673 from local dynamics}\label{12673local}

\begin{figure}[ht]
    \centering
    \includegraphics[width=\linewidth]{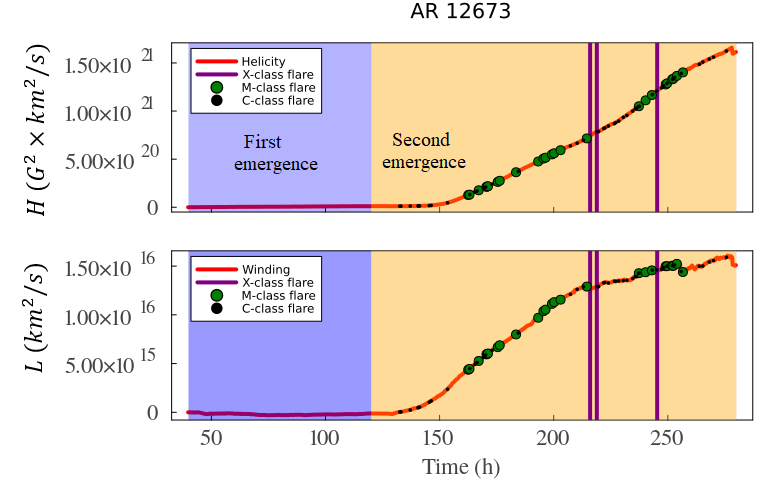}
    \caption{Accumulation of magnetic helicity (top) and magnetic winding (bottom) for AR 12673. Purple and beige backgrounds indicate first and second major emergences. Here, the start time is August 28, 2017, 9:00. 
    \label{helicity12673}}
\end{figure}

\begin{figure}[ht]
    \centering
    \includegraphics[width=\linewidth]{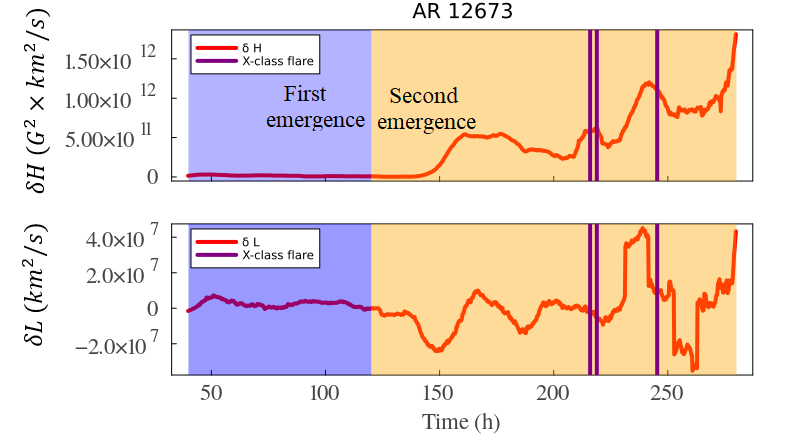}
    \caption{Evolution of the magnetic helicity imabalance (top) and magnetic winding imbalance (bottom) for AR 12673. Negative values indicate the injection of potential-dominated magnetic structure, positive values indicate dominance of current-carrying dominated magnetic field structure.Purple and beige backgrounds indicate first and second major emergences.
    \label{imbalance12673}}
\end{figure}

Several studies analysed the topology of the magnetic fields of AR 12673 in terms of its magnetic helicity. \citet{vemareddy2019very} showed a rapid input of helicity before the occurrence of the major flares, while \citet{moraitis2019magnetic} highlighted the role of the current-carrying component of the helicity. More recently, \cite{raphaldini2022magnetic}, computed the magnetic winding, a direct measure of the topology of the magnetic field configuration, to show that, compared with other active regions, the complexity of the magnetic field  of this region is much more dominated by current-carrying structures.

Figure \ref{maghel} shows the evolution of the magnetic field and helicity densities throughout the AR 12673 lifetime. Up to time t=100 hours after since the appearance of AR 12673 on disc both the magnetic fields (see \ref{maghel} (a) and (c) ) and the helicity (see \ref{maghel} (b) and (d)) are dominated by a single spot, reminiscent of AR 12665 and 12670  during previous Carrington rotations. It is also notable that at time $t$=100 hours the helicity structure is weaker than it was at $t$=10 hours, suggesting a decay or diffusion of helicity throughout this timespan; on the other hand the magnetic field structure displays a magnetic field that remains coherent. At this point new magnetic field structures start to emerge around the old spot, which remains relatively unperturbed by the new emergence. At $t$=150 hours (\ref{maghel} (g) and (h)) present an increased complexity in the pattern of magnetic fields and magnetic helicity, with the imprint of the old spot still present. At t=200 hours, close to the sequence of two intense X-class flares, including the X-9 flare, we see a remarkably complex magnetic field structure, characteristic of a delta spot, where the imprint of the old spot can still be identified. This interaction between different emergences at the same location clearly played a role in the complexity of the magnetic fields in AR 12673, raising some important questions: How do the large scale magnetic fields and flows at the depth where the emergences originate favour sequential rise of magnetic fields through the convection zone? What are the large scale patterns that favour these sequential emergences? How does the old decaying spot remain so stable despite of the strong perturbations resulting from the secondary emergence of AR 12673? 

The accumulated helicity and winding time series are illustrated in figure \ref{helicity12673} where it is shown that very little magnetic helicity and winding are accumulated while the old spot is the dominant structure, until about 120 hours after emergence. As soon as the second emergence takes place, a steady accumulation of helicity and winding with flare activity is initiated a few hours after. The flare activity is initially comprised of C-class flares, followed by a sequence of M-class flares at around 170 hours. As the helicity and winding increases, accompanied by several M-class flares two sequential X-class flares appear,separated by 3 hours, at around 220 hours, including the X-9.3 flare. The X-class flares are accompanied by a small inflection in both curves. About 20 hours after the X9.3 flare, another X-class flare was produced, this time a X-1.3 flare. Two days later another X-class flare, a X-8.3 flare, was produced, which is not presented here since it took place when the active region was near the border of the disk.

Next, we analyse the imbalance between the current-carrying and potential parts of the helicity and winding, also known as the $\delta$ quantities difined in eqs. (\ref{deltahel}) and (\ref{deltawin}) . Since these signals are very noisy we present a smoothed version of $\delta H'$ and $\delta L'$ over a 10 hour window . We present the evolution of the delta quantities for AR 12673 in fig.\ref{imbalance12673}. As expected, very little imbalance is observed in AR 12673 while it has an $\alpha$-type single spot structure. Once the second emergence takes place an imbalance develops in both the magnetic helicity and winding. We notice that the first response is seen in the winding imbalance, with negative values, indicating that in the early stages of the secondary emergence the magnetic field structures have low intensity and are primarily potential. Once the winding imbalance starts becoming positive the imbalance in the helicity also starts developing, now both indicating current-carrying dominance, the increase in both indicates that complex morphology is present in regions with strong magnetic fields. The difference between both curves can be understood from the fact that, while the helicity is weighted by magnetic field strength, the magnetic winding is not. The helicity imbalance in particular shows a clear pattern of increase before X-flare activity, followed by a decrease, probably due to dissipative relaxation due to the reconnection of magnetic field lines.

\subsection{A Solar Cycle Peak-phase Storm}

\begin{figure}[ht]
    \centering
    \includegraphics[width=0.85\linewidth]{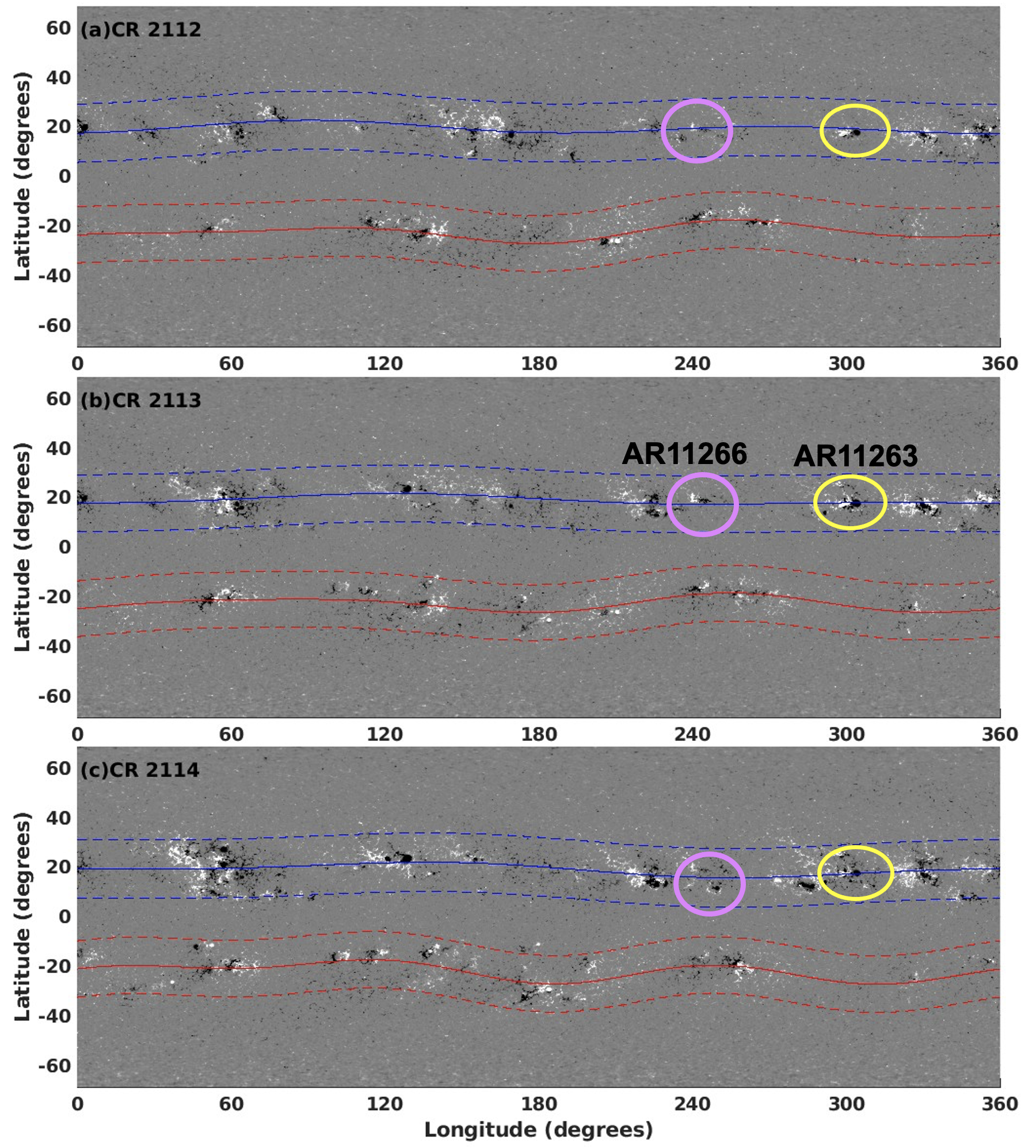}
    \caption{Evolution of magnetic toroid pattern during Carrington Rotations CR2112 (Jul 03-Jul 30, 2011), CR2113 (Jul 30-Aug 26, 2011) and CR2114 (Aug 26- Sep 22, 2011), i.e., from one CR before through the storm to one CR after, during which AR11263, located at 300-degree longitude in North (yellow-circled region), produce the second biggest flare (X6.9) of cycle 24. Another active region (AR11266, marked by pink-circled region) in the same toroid, located at a longitude of about 55-degree left of AR11263, did not flare.
    \label{Peak-phase-storm}}
\end{figure}

Having derived the pre-solar-storm features of an active region that resulted in a strong flare activity during the solar minimum-phase, we now compare them with the features of a flare-producing active region, AR 11263, during the peak-phase of cycle 24.

AR11263 appeared on the east limb of the visible disk on the 28th of July of 2011, during the late rising phase of cycle 24. It is to be noted that cycle 24 exhibited widely separated double peaks -- the first peak occurred in 2011 in the North hemisphere, while the second peak was in 2014 in the South \citep{pesnell2016predictions}. AR 11263 occurred near the first peak in the then dominant northern hemisphere. It appeared on the disk already with a $\beta\gamma\delta$ configuration, producing a C-class flare within a day after its appearance. It kept a moderate level of activity until the active region evolved into a more complex structure, and a consequent increase in the level of activity around the 3rd of August, resulting in an M-class flare. On August 8 a second emergence took place, further increasing the level of complexity of the active region, leading to a sharp increase in the level of activity, resulting in two M-class flares and one X class-flare (X-6.9), the third largest flare of solar cycle 24. After that, the level of activity decreased, with the active region presenting a few C-class flares before it moved to the limb.

What can we analyse with one CR lead time from the global toroid pattern? In the following two subsections \ref{secAR11263global}  and \ref{AR11263local} respectively, we present results from the global magnetic field configurations during Carrington rotations 2112-2114 and the magnetic helicity and winding input of AR 11263, as well as current carrying/potential decompositions, several hours before the X-flare.

\subsubsection{Analyses of pre-storm features of AR11263 from global toroid patterns} \label{secAR11263global}

Three panels of Figure \ref{Peak-phase-storm} display the evolution of toroid patterns, derived from the surface distribution of active regions, during CR2112-2114. AR11263 is located in the yellow-circled region, at a longitude of $300^{\circ}$ in the North. In the same toroid we mark another non-flaring active region, AR11266, by a pink circle, located at about $245^{\circ}$ longitude, to compare the pre-storm features of a flaring and a non-flaring active region. 

Several features are immediately revealed from these three panels. First, both the North and South toroids show sustaining multi-mode patterns, with North toroid dominated by $m=1$ longitudinal mode and the South by $m=1$, 2 and 3 modes. The North hemisphere magnetic activity being much stronger than the South during 2011, it is expected   since the $m=1$ mode is a characteristic of the tipping of a toroidal ring that behaves rigid like a "steel ring" \citep{cally2003clamshell}. \citet{cally2003clamshell} showed that a toroidal ring tips with an $m=1$ pattern, when the magnetic field is strong enough, typically more than 30 kGauss, whereas the weak rings with field strengths smaller than 30 kGauss deform, and reveal a pattern dominated by $m>1$ modes. Thus a dominant $m=1$ pattern in the North during 2011 storm by AR 11263 reveals the stronger toroidal band in the North than in the South. Second, AR11263 appears in the tipped-away portion of the North and South toroids, whereas AR11266 appears in a latitude-longitude location where the North and South toroids are closer to each other, thus allowing strengthening of AR11263 and weakening of AR11266 due to interactions between oppositely-directed toroids. The global patterns here are much closer to those during the Halloween storms of cycle 23, except that cycle 24 was altogether much weaker than cycle 23. The pre-storm features from CR2112 indicate the possibility of upcoming big storms from AR11263 within one or two CRs, and perhaps much less possibility of storms occurring from AR11266.

A comparison of Figure \ref{Peak-phase-storm} with Figure \ref{Minima-phase-storm} clearly reveals relatively much more sustained patterns in the evolution of toroid during the peak-phase with respect to that during the minima-phase of the cycle. This is related to the deeper origin of these warped toroid patterns, which are most likely produced by the interaction of Rossby waves with the dynamo-generated toroidal magnetic fields. Those Rossby waves were found to have variation in their drift-speed in longitude as function of latitude. Figure 7 of \citet{Dikpati_2018} shows that the phase-speed is close to zero when the toroidal band is around $20^{\circ}$ latitude, which is the case here for the peak-phase toroid patterns during CR2112-2114. The more sustained these patterns are the more flare-prone the big and complex active regions are from the tipped-away portion of the toroids.

While the peak-phase storms are more predictable, with a significant lead time, from the analysis of global toroid patterns, the obvious question is: what additional pre-storm features can we find from the local dynamics several hours ahead of the storm?

\clearpage

\subsubsection{Analysing pre-storm features of AR11263 from local dynamics}\label{AR11263local}
\begin{figure}[ht]
    \centering
    \includegraphics[width=0.8\linewidth]{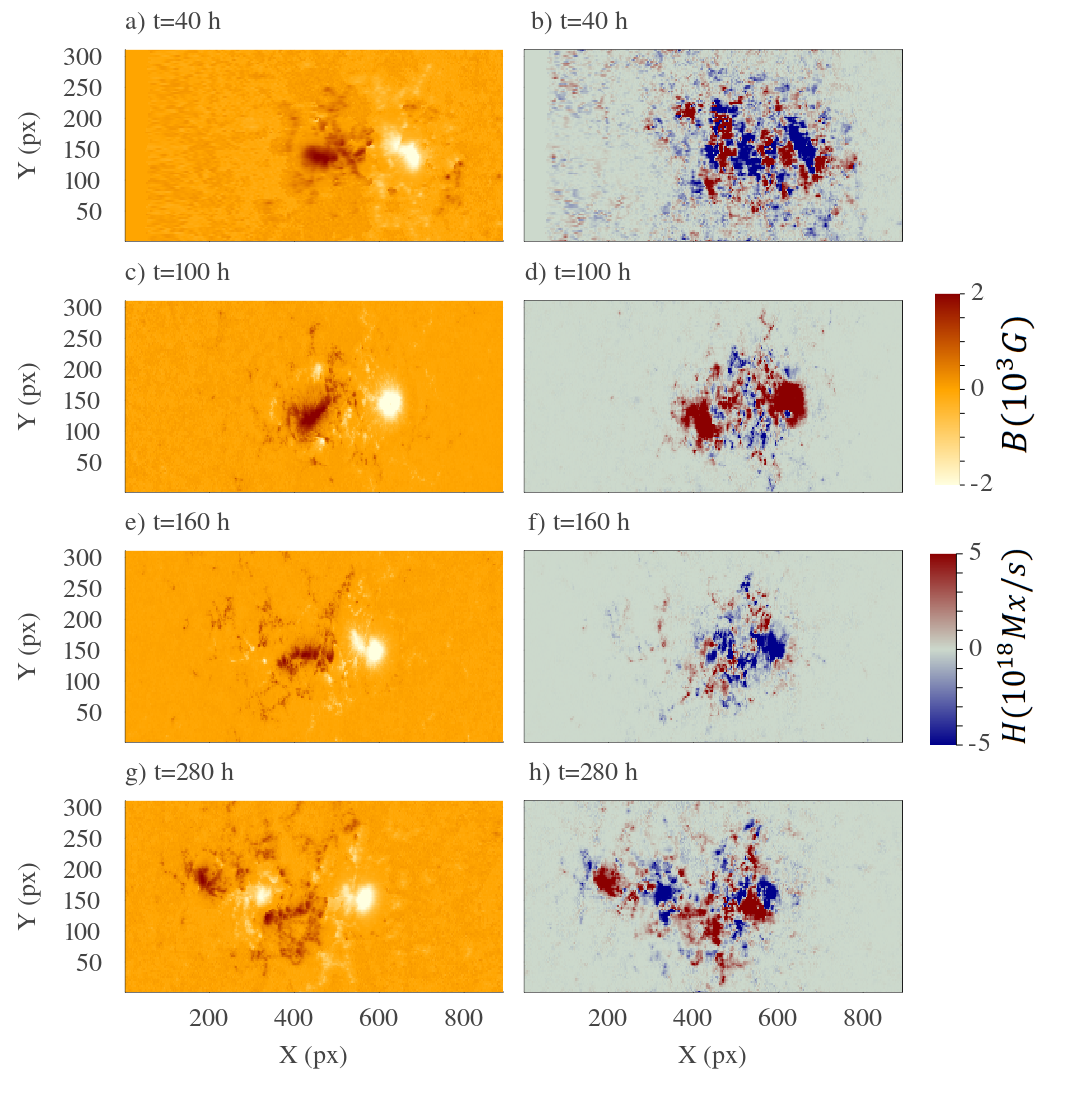}    \caption{Evolution of the magnetic fields (left) and helicity flux helicity  (right) for AR 11263. Time evolution movies of AR 11263 can be seen  \href{https://zenodo.org/record/8305952}{in continuum intensity and magnetogram.}
    \label{AR11263FIELDS}}
\end{figure}

\begin{figure}[ht]
    \centering
    \includegraphics[width=\linewidth]{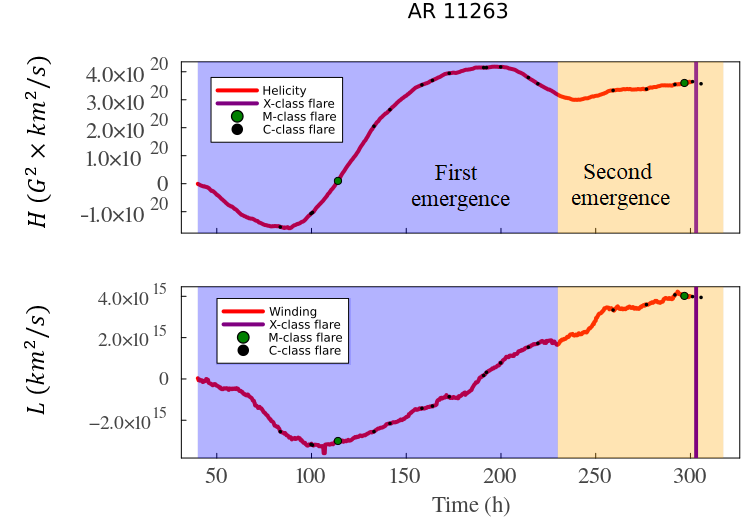}
    \caption{Accumulation of magnetic helicity (top) and magnetic winding (bottom) for AR 11263.Purple and beige backgrounds indicate first and second major emergences. We note that after the X-class flare the AR approaches the limb, therefore the data becomes less reliable and is omitted.Here the start time is July 27, 2011, 12:00 hrs. 
    \label{AR11263N}}
\end{figure}

\begin{figure}[ht]
    \centering
    \includegraphics[width=\linewidth]{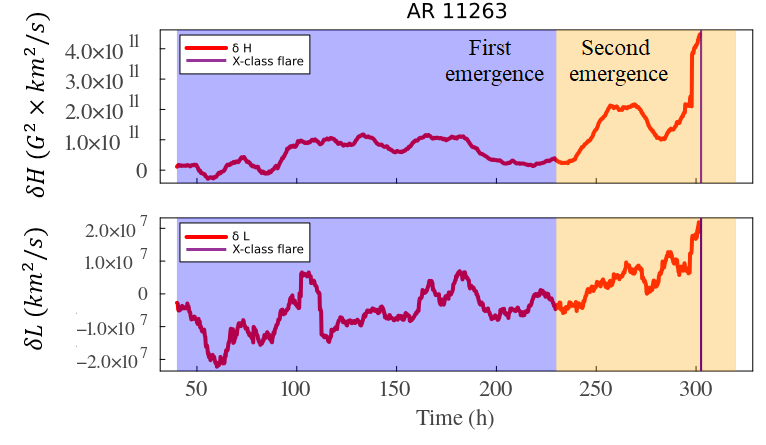}
    \caption{Evolution of the magnetic helicity imabalance (top) and magnetic winding imbalance (bottom) for AR 11263. Negative values indicate the injection of potential-dominated magnetic structure, positive values indicate dominance of current-carrying dominated magnetic field structure. Purple and beige backgrounds indicate first and second major emergences.
    \label{IMBALANCE11263}}
\end{figure}

The evolution of the magnetic field morphology in AR 11263 is described in Figure \ref{AR11263FIELDS}. There we see that the active regions appeared on disk, first as a simple $\beta$-type spot, until at around $t=200$ hours a second emergence interacted with the $\beta$ configuration resulting in a complex magnetic field configuration, a $\beta-\gamma-\delta$-type structure. The evolution of the magnetic field and helicity densities is depicted in Fig. \ref{AR11263FIELDS}.

The analysis of the accumulated magnetic helicity and winding for this active region is presented in figure \ref{AR11263N}. Initially AR 11263 showed negative helicity accumulation which was reversed by added positive helicity starting at around $85$ hours, which persisted until roughly $t=200$ hours, followed by a small decrease in the total winding. As soon as the second emergence took place, a slow but steady increase in the helicity took place until the time of the X 6.9 flare. In terms of the accumulation of magnetic winding, the overall trend was similar, except that the winding presented an almost steady increase from  $t=100$ hours until the time of the X flare at around $t=300$ hours. A sudden relative rapid increase in the winding curve was observed following the second emergence at around $t=250$ hours, which became more evident about 10 hours before the flare happened.

Results from the magnetic helicity and magnetic winding decompositions into potential and current/carrying fields is presented in Figure \ref{IMBALANCE11263}.  The helicity and winding imbalances (or $\delta$ measure)  for the magnetic helicity imply an overall dominance of potential structures, followed in time by a stable input of current-carrying-dominated structures from $t=85$ hours until time $t=180$ hours, and after that, a decrease in $\delta H'$ that indicates a balance between potential and current carrying helicity between $t=200$ and $t=250$ hours. Starting at time $t=250$, two impulses in $\delta H'$, the second and largest ones leading up to the X-class flare. The magnetic winding imbalance, on the other hand, showed a fluctuating behavior with a predominance of potential fields after the second emergence, indicating that in the beginning the secondary emergence was characterized by weak potential fields, similar to what is observed in AR 12673. At around  $t=250$ there was a steady increase in this quantity, followed by a sharp increase just before the flare. Again, as in the case of AR 12673, the magnetic helicity and winding imbalance shows clear a tendency of increasing before the major flare, which is more evident for the magnetic helicity, confirming the role of the injection of current-carrying dominated structures, leading up to the X-class flare events.

\begin{figure}[ht]
    \centering
    \includegraphics[width=\linewidth]{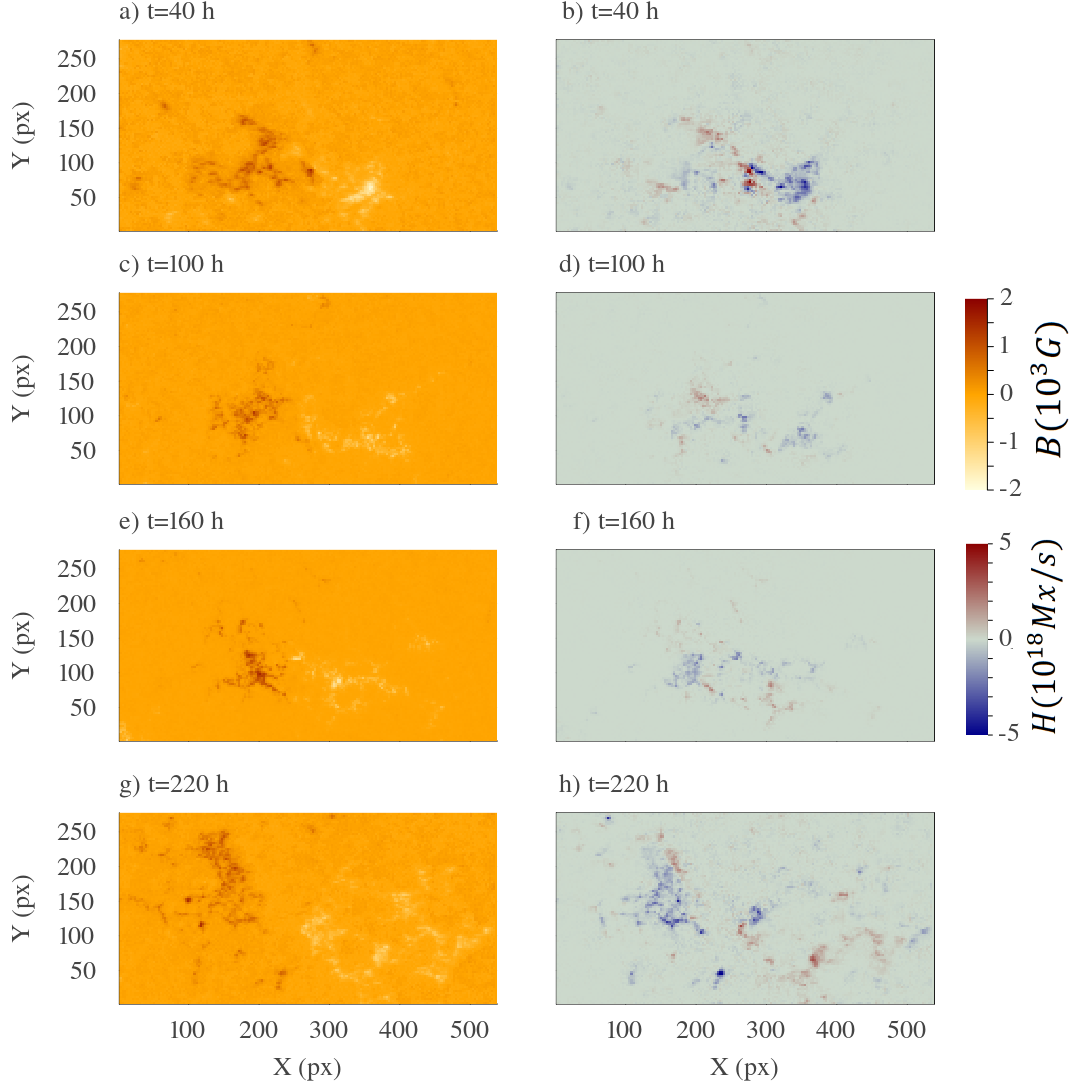}
    \caption{Evolution of the magnetic fields (left) and helicity \textcolor{red}{flux} (right) for AR 11266. Time evolution movies of AR 11263 can be seen \href{https://zenodo.org/record/8305952}{in continuum intensity and magnetogram}.
    \label{AR11266FIELDS}}
\end{figure}

\begin{figure}[ht]
    \centering
    \includegraphics[width=0.9\linewidth]{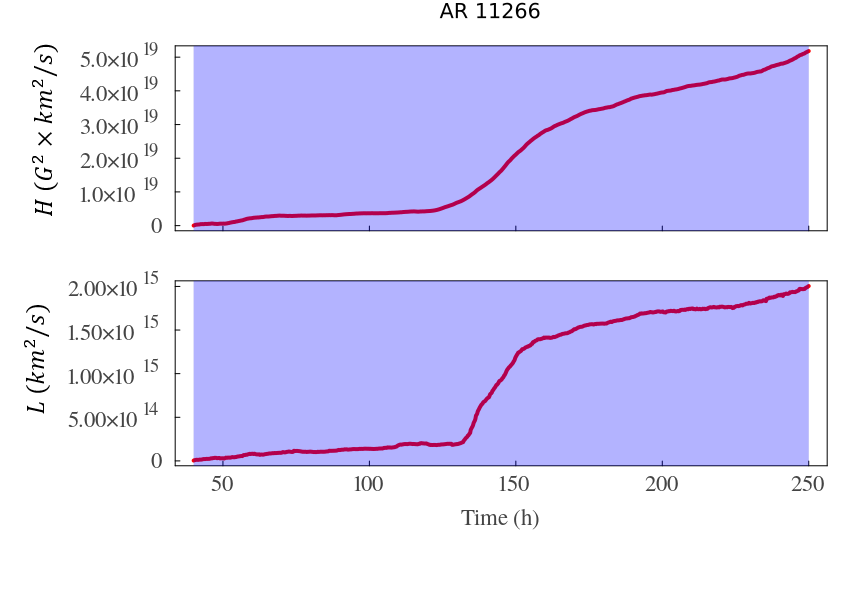}
    \caption{Accumulation of \textcolor{red}{fluxes} magnetic helicity (top) and magnetic winding (bottom) for AR 11266. Here, the start time is August 2, 2011, 10:00 hrs. 
    \label{AR11266N}}
\end{figure}

\begin{figure}[ht]
    \centering
    \includegraphics[width=0.9\linewidth]{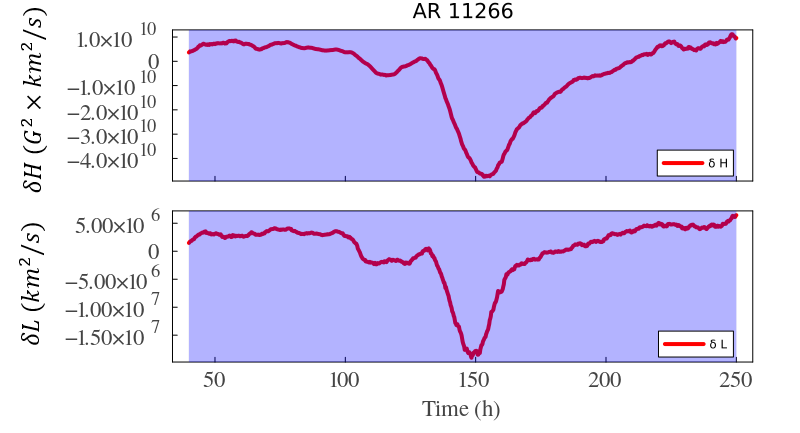}
    \caption{Evolution of the magnetic helicity imabalance (top) and magnetic winding imbalance (bottom) for AR 11266. Negative values indicate the injection of potential-dominated magnetic structure, positive values indicate dominance of current-carrying dominated magnetic field structure.
    \label{IMBALANCE11266}}
\end{figure}

\subsection{A peak-phase active region, AR 11266, triggering no storm}\label{AR11266local}
Active region AR 11266 emerged on the disk on August 1st, 2011, just a few days after AR 11263; it was the first active region to emerge on the northern hemisphere following AR 11263. Throughout its transit across the disk it was characterized by a $\beta$-type structure with well defined positive and negative polarities. The active regions started losing coherence at around 180 hours. Figure \ref{AR11266FIELDS} shows the evolution of the magnetic fields in AR 11266; we can see that the magnetic field morphology in AR 11266 starts losing coherence after about $t=200$ hours. So, unlike the in the cases of AR 12673 and AR 11263 analysed in subsections \ref{12673local} and \ref{AR11263local}, the simple initial configuration remained until it decayed, instead of interacting with a new emergence.

Figure \ref{AR11266N} shows the evolution of the accumulated total magnetic helicity and winding. In terms of the accumulation of magnetic helicity and winding, both quantities show a very similar behavior with a slow increase up to $t=130$ hours, followed by a rapid impulse until $t=160$ hours, after which the accumulation of both quantities slowed down.
The inspection of $\delta H'$ and $\delta L'$ reveal the nature of the rapid increase in the accumulated helicity and winding around $t=150$ hours. Both the winding and helicity imbalances, shown in figure \ref{IMBALANCE11266}, demonstrate a clear dominance of potential structures during this time. Unlike AR 12673 and AR 11263, this active region did not present any major event of injection of current-carrying dominated structures, which explains why it did not flare.

\clearpage

\section{Summary}

Here, we summarize our main findings:

\begin{itemize}
  \item Pre-storm features were studied both in terms of the global distribution of active regions and individual configurations.
  
  \item Techniques based on individual active region's magnetic field characteristics provide several hours \citep{kusano2020physics,leka2019comparison} of lead time before major X-class flare events occur, leaving a short time for society to prepare against their hazardous effects. The identification of particular characteristics of the global magnetic field structure can potentially provide one Carrington rotation of lead time before big storms.

  \item The two most flare prolific active regions of solar cycle 24 were analysed: AR 12673, which occurred approaching the solar minimum, and AR 11263 which occurred near the maximum.

  \item Two classes of techniques were used to investigate pre-storm conditions: toroid tipping patterns were fitted to the global distribution of active regions, and AR's individual magnetic field morphologies were studied using the topological measures of magnetic helicity and winding.

  \item  AR 12673 emerged in the declining phase of the solar cycle 24, in September of 2017, when the solar photosphere wasn't populated by many active regions. As a result, the toroid fitting procedure presented in subsection \ref{AR11263global} showed a magnetic field band dominated by $m=1$ wavenumber, in such a way that a tipping pattern could not be well determined. 
  
  \item The strong flare activity of AR 12673 resulted from the interaction between an old decaying spot and a new emergence, producing a highly complex magnetic field configuration. The local analysis showed that this interaction between an old spot and a new emergence resulted in a highly non-potential structure leading up to major X-class flares.
  
  \item For AR 11263, the toroid fitting procedure performed in subsection \label{AR11263global} revealed a scenario that is similar to that of the Halloween storm of 2003 \citep{dikpati2021deciphering}. In particular, AR 11263 appeared in a portion of the toroid where the north and southern hemisphere belts tip away from each other. 
  
  \item Like AR 12673, the X-class flare produced by AR 11263 occurred after a new emergence took place at the same location where the $\beta$ structure was situated, resulting in a structure dominated by nonpotential field.
  
  \item AR 11266 was an active region adjacent to AR 11263. AR 11266 appeared as a  $\beta$-type structure that remained as such until it started diffusing out, towards the latter stages of its passage through the disk. The toroid fitting analysis shows that, unlike AR 11263, it was situated in a longitude where north and south hemisphere toroids come close to each other. And, unlike the case of AR 11263 and AR 12673, the emergence of AR 11266 was dominated by potential structures, which help us understand why it did not produce any flares at all.

  \item Comparison of figures \ref{imbalance12673}, \ref{IMBALANCE11263} and \ref{IMBALANCE11266}, with \ref{helicity12673}, \ref{AR11263N} and \ref{AR11266N} suggests that the magnetic helicity and winding imbalance quantities are a better predictor than the respective raw quantities. This highlights the importance of relative contribution of the current-carrying heliciy/winding.
\end{itemize}

Figure \ref{sum} and Table \ref{Table} summarize the findings in terms of the maximum imbalance found in the helicity/winding analysis, representing either the main emergence event dominated by potential (negative) values of current carrying  (positive) structures. 




\begin{center}\label{Table}
\begin{tabular}{||c c c c c||} 
 \hline
 AR  & H & L & $ \delta $H$$ & $ \delta $L$$ \\ [1ex] 
 \hline\hline
 11266 & 5.1$\times 10^{19}$ & 2.1$\times 10^{15}$ &-4.6$\times10^{11}$& -1.8$\times 10^{7}$ \\ 
 \hline
 11263 & 4.2$\times 10^{20} $ & 4.1$\times 10^{15}$ &4.4$\times 10^{11}$ &2.2$\times 10^{7}$ \\
 \hline
 12673 &1.8 $\times 10^{21}$& 1.5 $\times 10^{17}$ & 1.5$\times10^{12} $& 4.3$\times10^{7}$\\
 \hline
\end{tabular}
\end{center}

\begin{figure}[ht]\label{sum}
    \centering
    \includegraphics[width=.9\linewidth]{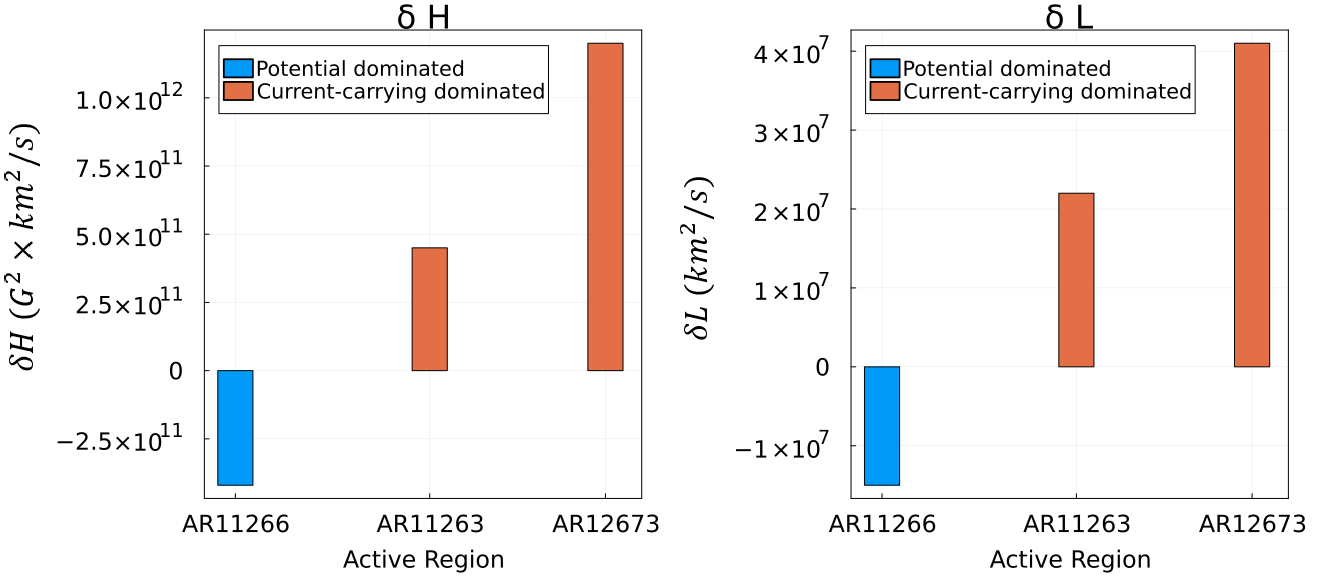}
    \caption{Summary of the maximal deviation of the helicity imbalance (right) and the magnetic winding imbalance (right).
    \label{sum}}
\end{figure}

\section{Concluding Remarks}

Forecasting the upcoming enhanced bursts of solar activity well ahead of time no doubt has significant societal value. The majority of energetic flares and coronal mass ejections that can adversely impact the techonological infrastructure occur during such bursty "seasons" of activity. So far primary emphasis has been given to forecasts of solar energetic phenomena occurring on hours-to-days time-scales and the configuration and evolution of individual active regions that are responsible for causing the events. These studies are important, and can have a lead time of a few to several hours to send out an alert. However, often predicting an upcoming solar storm with only hours lead time may not be enough to prevent the society from their hazardous impacts. 

In a recent study by \citet{dikpati2021deciphering}, in the context of analyzing the pre-solar-storm features of the 2003 Halloween storm, which caused multiple X-class flares, including an X-45 flare \citep{thomson2004ionosphere}, latitude-longitude location of active regions in a warped toroid and their evolution were found to play crucial roles in predicting the possibility of big storms. Using an MHD shallow-water model, which allows nonlinear interactions among Rossby waves, spot-producing toroidal fields at the base of the convection zone or tachocline and differential rotation there, the occurrence of Halloween storms from the same longitudes in both hemispheres within an interval of a few days was explained to be originating from the tipped-away portions of the warped toroids. When the toroidal bands coincide with the bulgings or high-pressure regions of the top surface of the shallow-water model, the flux can be pushed to the convection zone for buoyant rise to the surface. In the case of Halloween storms of 2003, toroid patterns were indicating the 'sympathetic emergences' of active regions at the same longitude in both hemispheres, and the toroid-patterns were evolving very slowly, revealing a sustained tachocline top-surface bulging. In such situations global toroid patterns and their evolutions can be used to forecast the possibility of upcoming big storms. 

In the present work, we examined whether we can forecast the biggest storm of cycle 24, namely the 2017-September storm, by studying the evolutionary patterns of global toroid. By using an optimization technique based on Trust Region Reflective (TRR) algorithm, we derived the toroid patterns for three CRs covering the span before and after the storm. We found that the patterns were strongly dominated by $m=1$ mode in longitude. Noting that 2017-September storm occurred during the minima-phase of cycle 24, we can expect $m=1$ toroid-patterns, because only a few active regions would emerge at that cycle-phase. Even though toroid patterns indicated 'sympathetic emergences' in both hemisphere, but North and South hemispheres' active regions were not from the 'tipped-away' portions of the the North and South toroids, instead they were in close proximity. Thus they are expected to weaken each other, ruining the possibility of big storms. However, due to new emergence at the center of the apparently inactive decaying AR12673 in the South, that active region became extremely active due to the complex interactions among old and new active regions, and caused the biggest storm, including four X-class flares (the biggest one being X-9.3, the biggest of cycle 24) and 27 M-class flares. 

While global toroid patterns can explain why new emergences occurred at the same longitude, namely due to sustained bulging of the shallow-water tachocline top-surface containing toroidal band, we need additional physics to understand this solar minima-phase storm, through the analyses of the configuration and evolution of AR 12673 as function of time. Thus, by analysing the measures of magnetic field topology/complexity, namely the magnetic helicity and the magnetic winding, we can understand that the rate of huge build-up of the current-carrying structure in the configuration of AR 12673 within about 100 hours from the new emergence at the center of decaying AR 12673 was the major cause of 2017-September solar storm.

In general the minima-phase solar storms are bound to be unusual, due to originating from the ARs that do not strictly follow a sustained toroid patterns. Hence, unlike the Halloween storms of 2003, the 2017-September storm cannot be speculated with one CR lead time just from the analyses of the global toroid patterns. Additional physics from the local configuration and evolution of the active region responsible for causing the storms was necessary to speculate this storm, with the lead time reduced to several hours from several days. 

On the other hand, the second biggest storm of cycle 24 occurred during August 2011, during the peak-phase of that cycle. So, very much like the Halloween storms occuring during the peak-phase of cycle 23, we showed that the pre-storm features of slowly evolving magnetic toroid patterns can anticipate the 2011-August storm originating from the AR 11263; this storm occurred from an AR in the tipped-away portions of the North and South toroids. This forecast is possible because many active regions' emerge during the peak-phase, and lead to the toroid patterns with multiple $m$ modes. Furthermore, the toroid patterns evolve slowly, because the speed of the magnetically modified Rossby waves is very slow at the latitude of $\sim 15-20^{\circ}$ where the toroids are located during the peak-phase of a cycle. Sympathetic emergences can occur too due to sustained bulging. 

What can the local dynamics add to the pre-storm features for predicting storms in this case? The magnetic helicity and winding analysis for AR 11263 indicated a substantial injection of current-carrying structure, even though one order of magnitude less than that in AR 12673, and eventually caused the occurrence of the second biggest flare (of class X-6.9) of cycle 24.

Around the time of AR 11263 the Sun was populated with many sunspots. In order to understand how the latitude-longitude location on the warped toroid can be indicative of pre-storm properties of the active region, we analysed another active region AR 11266, which emerged about 45-degrees away in longitude compared to the neighboring AR 11263. Analyses of toroid patterns indicated that AR 11266 was located in the portion of the toroid, which was in close proximity of its opposite hemisphere counterpart, and hence got weakened due to the interaction between them. Obviously big storms are not anticipated from this active region from the global toroid analysis. However, during the lifetime of this AR complexity may arise from new emergences at the same longitude; hence the study of local dynamics is necessary. The magnetic helicity and winding computation showed that this AR was neither associated with a sympathetic emergence nor a current-carrying structure developed during its life time. Thus this AR did not produce flare, confirming the pre-storm conjectures we derived from both global and local dynamics.

We conclude, from the main findings from the analysis of pre-storm properties, that global toroid patterns and their slow evolution during peak-phase can lead to major flare activity from those ARs which emerge in a longitudinal position where north and south hemisphere toroids tipped away from each other. This gives enough lead time to follow how the AR configuration is locally evolving, namely whether a large helicity and magnetic winding are building up or not, and whether the current-carrying structure is getting built-up or just the potential-field structure is continuing. In the case of minima-phase storms, global toroid patterns evolve much faster than that in the peak-phase; hence a close look at the sympathetic emergence and their local dynamics build-up is the best reliable signal that can be sent to prevent the hazardous impact of the unusual big storms during the minima-phase, despite having only just hours' lead time. 

In a future study, we will be examining more peak-phase and minima-phase storms to compare and build better statistics. Combining the evolutionary patterns of the global toroids in which the active regions are tightly stringed, along with the evolution of the local configuration of individual, flare-prolific active regions, the storms can be predicted with a significant lead time of several days. A natural question that should be answered is to whether the relationship between tipped-away regions of the toroid and strong helicity/winding input always holds or if such active regions can also exist in longitudes where north/south toroids approach each other. 

The analysis and derivation of pre-solar-storm features we presented here is very similar to the Earth's weather prediction from the troughs and ridges associated with Rossby waves, which provide the conditions for the formation of storms \citep{holton1973introduction}. The analysis of atmospheric synoptic charts \citep{bluestein1992synoptic} allowed meteorologists to predict the weather with the lead time of a few days.  Similarly, our analysis suggest the solar storms can be predicted with one solar rotation lead time, meaning about one month lead time, if we can derive the pre-storm features utilising the information from both the global toroid patterns in which flare-prone active regions manifest and their local dynamics.

\section{acknowledgments*}
This work is supported by the National Center for Atmospheric Research, which
is a major facility sponsored by the National Science Foundation under
cooperative agreement 1852977. We acknowledge support from several NASA grants, 
namely MD and BR acknowledge NASA-LWS award 80NSSC20K0355, NASA-HSR award 
80NSSC21K1676. MD also acknowledges COFFIES Phase II NASA-DRIVE Center for the subaward from Stanford with award number 80NSSC22M0162. AN acknowledges NASA DRIVE Center COFFIES grant 80NSSC20K0602. ASWT has been supported by Fundação de Amparo à Pesquisa do
Estado de São Paulo (FAPESP) (grant 2020/14162-6).


\section*{Appendinx: Potential and current-carrying helicity and winding}
Several works \citep{pariat2017,raphaldini2022magnetic} have highlighted the importance not only of the total heliticy for the understanding of the flaring activity in active regions, but also, their decompositions in terms of potential and current-carrying components. If we define the current-carrying magnetic field as $\textbf{B}_C=\textbf{B}-\textbf{B}_P$. Then the following decompositions can be defined for the magnetic field foot-point motion:

\begin{equation}
    \textbf{u}(\textbf{x})=\textbf{v}_{\parallel}(\textbf{x})-\frac{v_z}{B_z}\textbf{B}_{\parallel c}(\textbf{x})-\frac{v_z}{B_z}\textbf{B}_{\parallel p}(\textbf{x})
\end{equation}

where $\textbf{v}_{\parallel}(\textbf{x})$ is the in plane component of the velocity, also known as "braiding-motion", the current-carrying velocity is then defined as:

\begin{equation}
   \textbf{u}_c(\textbf{x})= \textbf{v}_{\parallel}(\textbf{x})-\frac{v_z}{B_z}\textbf{B}_{\parallel c}(\textbf{x})
\end{equation}

while the potential component of the velocity is defined as:

\begin{equation}
   \textbf{u}_p(\textbf{x})= \textbf{v}_{\parallel}(\textbf{x})-\frac{v_z}{B_z}\textbf{B}_{\parallel p}(\textbf{x})
\end{equation}

where the symbol $\parallel$ detones the parallel component of the respective quantity. Finally, we define the current-carrying component of the helicity as
\begin{equation}
    H_c=\frac{1}{2\pi}\int_0^T\int_{\partial V}\int_{\partial V}B_z(\textbf{x})B_z(\textbf{y})\frac{(\textbf{u}_c(\textbf{x})-\textbf{u}_c(\textbf{y}))\times(\textbf{x}-\textbf{y})}{\vert \textbf{x}-\textbf{y}\vert ^2} {\rm d}^2\textbf{y} {\rm d}^2\textbf{x} dt
\end{equation}

and the potential component of the helicity as:

\begin{equation}
    H_p=\frac{1}{2\pi}\int_0^T\int_{\partial V}\int_{\partial V}B_z(\textbf{x})B_z(\textbf{y})\frac{(\textbf{u}_p(\textbf{x})-\textbf{u}_p(\textbf{y}))\times(\textbf{x}-\textbf{y})}{\vert \textbf{x}-\textbf{y}\vert ^2} {\rm d}^2\textbf{y} {\rm d}^2\textbf{y} dt
\end{equation}
Respective densities of injection of current-carrrying and potential helicity are defined as:

\begin{equation}
\mathcal{H}_c(\textbf{x})=\int_{\partial V}B_z(\textbf{x})B_z(\textbf{y})\frac{(\textbf{u}_c(\textbf{x})-\textbf{u}_c(\textbf{y}))\times(\textbf{x}-\textbf{y})}{\vert \textbf{x}-\textbf{y}\vert ^2} {\rm d}^2\textbf{y} 
\end{equation}

and,

\begin{equation}
    \mathcal{H}_p(\textbf{x})=\int_{\partial V}B_z(\textbf{x})B_z(\textbf{y})\frac{(\textbf{u}_p(\textbf{x})-\textbf{u}_p(\textbf{y}))\times(\textbf{x}-\textbf{y})}{\vert \textbf{x}-\textbf{y}\vert ^2} {\rm d}^2\textbf{y} 
\end{equation}

Similar definitions hold for the magnetic winding, we define the current-carrying component of the winding as
\begin{equation}
    L_c=\frac{1}{2\pi}\int_0^T\int_{\partial V}\int_{\partial V}I_z(\textbf{x})I_z(\textbf{y})\frac{(\textbf{u}_c(\textbf{x})-\textbf{u}_c(\textbf{y}))\times(\textbf{x}-\textbf{y})}{\vert \textbf{x}-\textbf{y}\vert ^2} {\rm d}^2\textbf{y}  {\rm d}^2\textbf{x}  dt
\end{equation}

and the potential component of the winding as:

\begin{equation}
    L_p=\frac{1}{2\pi}\int_0^T\int_{\partial V}\int_{\partial V}I_z(\textbf{x})I_z(\textbf{y})\frac{(\textbf{u}_p(\textbf{x})-\textbf{u}_p(\textbf{y}))\times(\textbf{x}-\textbf{y})}{\vert \textbf{x}-\textbf{y}\vert ^2} {\rm d}^2\textbf{y}  {\rm d}^2\textbf{x} dt
\end{equation}
Respective densities of injection of current-carrrying and potential winding are defined as:

\begin{equation}
    \mathcal{L}_c(\textbf{x})=\frac{1}{2\pi}\int_{\partial V}I_z(\textbf{x})I_z(\textbf{y})\frac{(\textbf{u}_c(\textbf{x})-\textbf{u}_c(\textbf{y}))\times(\textbf{x}-\textbf{y})}{\vert \textbf{x}-\textbf{y}\vert ^2} {\rm d}^2\textbf{y} 
\end{equation}

and,

\begin{equation}
    \mathcal{L}_p(\textbf{x})=\frac{1}{2\pi}\int_{\partial V}I_z(\textbf{x})I_z(\textbf{y})\frac{(\textbf{u}_p(\textbf{x})-\textbf{u}_p(\textbf{y}))\times(\textbf{x}-\textbf{y})}{\vert \textbf{x}-\textbf{y}\vert ^2} {\rm d}^2\textbf{y}
\end{equation}

\bibliography{paper_2017storm}{}
\bibliographystyle{aasjournal}



\end{document}